\documentclass[preprint]{aastex631}
\usepackage{verbatim, placeins}

\newcommand{\num}[1]{\textcolor{black}{#1}}

\newcommand{\minimize}{$\texttt{SciPy.optimize.minimize()}$}
\newcommand{\neldermead}{\texttt{Nelder-Mead}} 
\newcommand{\newtoncg}{\texttt{Newton-CG}}
\newcommand{\bfgs}{\texttt{BFGS}}
\newcommand{\sfit}{\texttt{SFit}}

\begin{document}

\title{An Alternate Method for Minimizing $\chi^2$}

\author{Jennifer C. Yee}
\affiliation{Center for Astrophysics $|$ Harvard \& Smithsonian, 60 Garden St.,Cambridge, MA 02138, USA}
\email{jyee@cfa.harvard.edu}
\author{Andrew P. Gould}
\affiliation{Max-Planck-Institute for Astronomy, K\"onigstuhl 17, 69117 Heidelberg, Germany}
\affiliation{Department of Astronomy, Ohio State University, 140 W. 18th Ave., Columbus, OH 43210, USA}
\email{gould.34@osu.edu}

\begin{abstract}
In this paper, we describe an algorithm and associated software package (\texttt{sfit\_minimize}) for maximizing the likelihood function of a set of parameters by minimizing $\chi^2$. The key element of this method is that the algorithm estimates the second derivative of the $\chi^2$ function using first derivatives of the function to be fitted. These same derivatives can also be used to calculate the uncertainties in each parameter. We test this algorithm against several standard minimization algorithms in \minimize\, by fitting point lens models to light curves from the 2018 Korea Microlensing Telescope Network event database. We show that for fitting microlensing events, \sfit\, works faster than the \neldermead\, simplex method and is more reliable than the \bfgs\, gradient method; we also find that the \newtoncg\, method is not effective for fitting microlensing events.
\end{abstract}

\section{Introduction}

Model optimization is a significant component of most quantitative analyses. Many analyses in the field of microlensing have long used on an optimization algorithm that relies on only first derivatives of the function being fit to the data. Because this feature is distinct from many of the most readily available optimization algorithms, we present the derivation of this algorithm, a Python implementation (\sfit), and evaluate the performance of \sfit\, against existing algorithms implemented in \minimize\, for the microlensing use case.

{\section{The Derivation}
\label{sec:der}}

Our method builds upon the discussion in ``$\chi^2$ and Linear Fits" \citep{Gould03}, which noted that the approach to non-linear models could be expanded beyond the scope of that work. Suppose that the function we want to minimize is a function $F(x)$ that is described by $n$ parameters $A_i$ (where we use $A_i $, instead of $a_i$, as a reminder that in the general
case, they are non-linear). Considering
the general (nonlinear) case, we can Taylor expand $\chi^2$, in terms of the $n$ parameters:
\begin{eqnarray}
\chi^2 & = &  \chi^2_0 + \sum_i {\partial \chi^2\over \partial A_i} A_i
+ {1\over2}\sum_{i,j} {\partial^2 \chi^2\over \partial A_i \partial A_j}A_i A_j\\
& = & \chi^2_0 + \sum_i D_i * A_i + \sum_{i,j} B_{ij} * A_i A_j  + ... \quad,
\end{eqnarray}
where
\begin{eqnarray}
D_i  & \equiv & {\partial \chi^2\over \partial A_i} \label{eqn:d}\\
B_{ij} & \equiv & (1/2){\partial^2 \chi^2\over \partial A_i\partial A_j}  \quad. \label{eqn:b}
\end{eqnarray}
Then,
\begin{equation}
\frac{\partial \chi^2}{\partial A_i} = -2\sum_k \frac{(y_k - F(x_k))}{\sigma^2_k}\frac{\partial F(x_k)}{\partial A_i}
\end{equation}
and
\begin{equation}
\frac{\partial^2 \chi^2}{\partial A_i \partial A_j} = -2\sum_k \left[
 \frac{1}{\sigma^2_k}\frac{\partial F(x_k)}{\partial A_i}\frac{\partial F(x_k)}{\partial A_j} +
  \frac{(y_k - F(x_k))}{\sigma^2_k}\frac{\partial^2 F(x_k)}{\partial A_i \partial A_j}
 \right] \quad .
\end{equation}
In the special case of a linear function, $F(x) = \sum_i a_i f_i(x)$
then
\begin{equation}
\frac{\partial F(x)}{\partial a_i} = f_i(x)
\quad \mathrm{and} \quad
\frac{\partial^2 F(x)}{\partial a_i\partial a_j} = 0,
\end{equation}
so the second term disappears, and we find that the solution (derived from
second derivative of $\chi^2$) can be expressed
in terms of products of the first derivatives of the general functional
form.  For the general case, we simply make the approximation that the second derivative term can be neglected; i.e.,
\begin{equation}
\frac{\partial^2 \chi^2}{\partial A_i \partial A_j} \approx -2\sum_k 
 \frac{1}{\sigma^2_k}\frac{\partial F(x_k)}{\partial A_i}\frac{\partial F(x_k)}{\partial A_j} \quad.
\end{equation}

Hence, there are three ways to generalize Newton's method (actually discovered by Simpson)
to multiple dimensions:
\begin{enumerate}
\item{Use only first derivatives of the $\chi^2$ function (which is what Simpson did in
   1-D), the so-called gradient method.}
\item{Taylor expand $\chi^2$ and truncate at the second term, then
   solve this (very inexact equation) exactly by inversion
   of the matrix of second derivatives (Hessian).}
\item{First generalize Simpson's idea that a 1-D function is well
   described by its first derivative (which can be solved
   exactly) to several dimensions (i.e., assume the function is well
   described by a tangent plane) and solve this exactly, as is done here.}
\end{enumerate}

Because first derivatives are more stable than second derivatives, this algorithm could potentially be significantly more stable for situations in which the derivatives are derived numerically.

\section{Implementation}

\subsection{General}

We have implemented the above algorithm in the $\texttt{sfit\_minimizer}$ package. The goal was to make the calling sequence similar to that of \minimize:
\begin{equation}
\texttt{result = sfit\_minimizer.minimize(my\_func, x0=initial\_guess)}
\end{equation}
where $\texttt{my\_func}$ is an object of the type $\texttt{sfit\_minimizer.SFitFunction()}$. The user defines either the model, $F(x_k)$, or the residual, $y_k - F(x_k)$ where the $y_k$ are the data, calculation (i.e., the method $\texttt{my\_func.model()}$ or $\texttt{my\_func.residuals()}$). The user also defines
 the partial derivatives of the function to be minimized (i.e., the method $\texttt{my\_func.df()}$), $\partial F(x_k) / \partial A_i$. The package includes a simple example ($\texttt{example\_00\_linear\_fit.py}$) for fitting a linear model to demonstrate this usage.

The $\texttt{sfit\_minimizer.SFitFunction()}$ class contains methods that use the partial derivative function to calculate the next step from the $D_i$ and $B_{ij}$ following the method in \citet{Gould03} for linear functions. That is, $D_i$ and $B_{ij}$ are calculated from Equations \ref{eqn:d} and \ref{eqn:b}, respectively. Then, the step size for each parameter, $\Delta_i$, is
\begin{equation}
\Delta_i = \sum_j C_{ij} D_j \quad \mathrm{where} \quad C \equiv B^{-1} \quad,
\end{equation}
which is returned by $\texttt{sfit\_minimizer.SFitFunction.get\_step()}$.
The new value of $A_i$ is calculated by $\texttt{sfit\_minimizer.minimize()}$ to be
\begin{equation}
A_i = A_{i, 0} + \epsilon \Delta_i \quad . 
\end{equation}
In  $\texttt{sfit\_minimizer.minimize()}$, the user has the option to specify the value of $\epsilon$ or to make use of an adaptive step size, which starts at $\epsilon = 0.001$ and becomes larger as the minimum is approached.

Ultimately, $\texttt{sfit\_minimizer.minimize()}$ returns an $\texttt{sfit\_minimizer.SFitResults()}$ object that contains attributes similar to the object returned by \minimize. These include the best-fit values of the parameters, $\texttt{x}$, and their uncertainties $\texttt{sigma}$ (i.e., $\sigma_i = \sqrt{C_{ii}}$). For the rest of this paper, we will refer to our algorithm as \sfit\, for brevity.

\subsection{Microlensing-specific}

A point lens microlensing model \citep{Paczynski86b}, $A$, is described by a minimum of three parameters: $t_0$, $u_0$, and $t_{\rm E}$ \citep[for the definitions of these parameters see, e.g.,][]{Gaudi12}. In addition, there are two flux parameters, $f_{{\rm S}, k}$ and $f_{{\rm B}, k}$, used to scale the model to each dataset, $k$, to obtain the model flux: $f_{{\rm mod}, k} = f_{{\rm S}, k} A + f_{{\rm B}, k}$. 

The \texttt{MulensModel} package \citep{MulensModel} implements functions that calculate such models and their $\chi^2$s and derivatives relative to data. The $\texttt{mm\_funcs.py}$ module contains microlensing-specific implementations that use \texttt{MulensModel}. The class \texttt{PointLensSFitFunction} takes a  \texttt{MulensModel.Event} object as an argument and can be used with $\texttt{sfit\_minimizer.sfit\_minimize.minimize()}$ to obtain the best-fitting model parameters. This usage is demonstrated in $\texttt{example\_01\_pspl\_fit.py}$ for fitting the three standard Paczy\'{n}ski parameters above. An example additionally including the finite source parameter, $\rho$, is given in $\texttt{example\_02\_fspl\_fit.py}$ (note fitting $\rho$ requires \texttt{MulensModel} v3 or higher).

$\texttt{mm\_funcs.py}$ also includes a convenience function, $\texttt{fit\_mulens\_event()}$, that will automatically perform the fitting given a  \texttt{MulensModel.Event} object.
 
\section{Performance Test}

To test the performance of \sfit, we use the package to fit point-source--point-lens models \citep{Paczynski86b} to a sample of microlensing events from the Korea Microlensing Telescope Network \citep[KMTNet;][]{Kim16_KMTNet}. For comparison, we also perform the fitting using the \neldermead\, \citep{NelderMead}, \newtoncg\, \citep{QuasiNewtonMethods}, and \bfgs\, \citep{QuasiNewtonMethods} algorithms in \minimize. The \neldermead\, algorithm is a simplex algorithm, so it only relies on evaluating the $\chi^2$. In contrast to our algorithm, the \newtoncg\, and \bfgs\, algorithms use the jacobian of the likelihood function for the minimization. In all cases, we set \texttt{tol = 1e-5}.

\subsection{Sample Selection}

We select  our sample from microlensing events discovered in 2018 by KMTNet (\citealt{KimKim18_EF}, \citealt{Kim18EF,Kim18_AF}). We use only ``clear" microlensing events with reported fit parameters. We eliminate any events that were flagged as anomalous in the 2018 AnomalyFinder search \citep[although possible finite source or ``buried" host events were left in the sample;][]{Gould22AF5,Jung22AF6}. These cuts left 1822 events in the sample. 

For this sample, we use the online, $I$-band, pySIS \citep{Albrow09} data  from the KMTNet website (\url{https://kmtnet.kasi.re.kr/ulens/}). KMTNet takes data from three different sites and has multiple, sometimes overlapping, fields of observations. We treat data from different sites and different fields as separate datasets. For each dataset, we calculate the mean sky background and standard deviation as well as the mean and standard deviation of the full-width-at-half-max (FWHM) for each observation. We eliminate points with sky background more than 1 standard deviation above the mean or FWHM more than 3 standard deviations above the mean. This removes a large fraction of the outliers from the data. We also remove any points with negative errorbars or NaN for either the fluxes or errorbars.

\subsection{Fitting Procedure}

To find the initial starting value for the fit, we start by performing a series of linear fits using the EventFinder method (\citealt{KimKim18_EF}, \citealt{Kim18EF}). This method performs a linear fit over a grid of three parameters. From the best-fit EventFinder grid point, we take $t_0$, the time of the peak of the event. We remove any events for which the difference between the EventFinder $t_0$ and the reported KMTNet $t_0$ is more than \num{20} days (\num{40} events). We also remove any events whose light curves appear to be flat or anomalous (\num{8} additional events).

For the remaining events, we test a grid of values: $u_{0, i} = [0.01, 0.3, 0.7, 1.0, 1.5]$ and $t_{{\rm E}, j}=[1., 3., 10., 20., 40.]$. We perform a linear fit to the flux parameters  ($f_{{\rm S}, k}, f_{{\rm B}, k}$) and choose the $(u_0, t_{\rm E})$ pair with the smallest $\chi^2$ as the starting point for our fits.  Then, we renormalize the errorbars of each dataset. The initial errorbar renormalization factor for each dataset is calculated as the factor required to make the $\chi^2/\mathrm{d.o.f.} = 1$.

We fit for the optimal model parameters using each algorithm. We use \texttt{MulensModel} \citep{MulensModel} to calculate the point-lens microlensing light curve model, its derivatives, and the jacobians.  For the three \minimize\, algorithms, we perform a linear fit to the flux parameters at each iteration using built in functions from \texttt{MulensModel} and use the fitting algorithm to optimize $t_0$, $u_0$, and $t_{\rm E}$. For \sfit, we include the flux parameters as parameters of the fit. These distinctions mirror how we expect the algorithms to be used in practice for light curve fitting.

We remove the \num{34} events with $t_0$ for the best-fitting model outside the observing season (\num{$8168 < \mathrm{HJD}^\prime < 8413$}). Our final sample has \num{1716} events.

To ensure a statistically robust comparison of results we renormalize the errorbars again so the $\chi^2/\mathrm{d.o.f.} = 1$ relative to the best-fitting model and repeat the fitting from the original starting point. This can be important in cases for which the initial starting point is relatively far from the true fit, e.g., cases with a true value of $t_{\rm E} > 40~\mathrm{days}$. This renormalization can change the relative weighting of individual datasets, which in turn can affect the preferred model.

\subsection{Results}

For the second set of fits, we calculated several metrics to evaluate the performance of each algorithm. First, for a given event, we compared the $\chi^2$ of the best-fit reported by each algorithm to the best (minimum) value reported out of the four fits. The results are given in Table \ref{tab:dchi2} for several values of $\Delta\chi^2$ classified by whether or not the algorithm reported that the fit was successful (``reported success").

Each fit may be classified in one of four ways:
\begin{itemize}
\item{True positives: algorithm reported success and found the minimum.}
\item{False positives: algorithm reported success, but did not find the minimum,}
\item{True negatives: algorithm reported failure and did not find the minimum,}
\item{False negatives: algorithm reported failure, but found the minimum.}
\end{itemize}
For the purpose of these comparisons, we consider the algorithm to have found the minimum (``succeeded") if \num{$\Delta\chi^2 < 1.0$}.

We also calculated the number of $\chi^2$ function evaluations required by each algorithm for fitting each event. The maximum  number allowed was 999; if the number of evaluations exceeded this, the fit was marked as a failure. Table \ref{tab:nfev} provides statistical summaries of this metric.

Table \ref{tab:dchi2} shows that the \sfit\, algorithm had the highest reliability for fitting microlensing events. It had very low false positive (\num{0\%}) and false negative (\num{1\%}) rates. The \bfgs\, and \neldermead\, algorithms both had low rates of false positives (\num{0\% and 1\%}, respectively) but most of the reported failures were false negatives (\num{$\sim 100\%$ and 91\%}, respectively). The false positives and false negatives for these two algorithms accounted for \num{32\% and 13\%} of the fits, respectively. For the \newtoncg\, algorithm, \num{36\%} of the reported successes were false positives and \num{22\%} of the reported failures were false negatives, accounting for \num{46\%} of the total fits. 

\begin{figure}
	\begin{centering}
	\includegraphics[height=0.5\textheight]{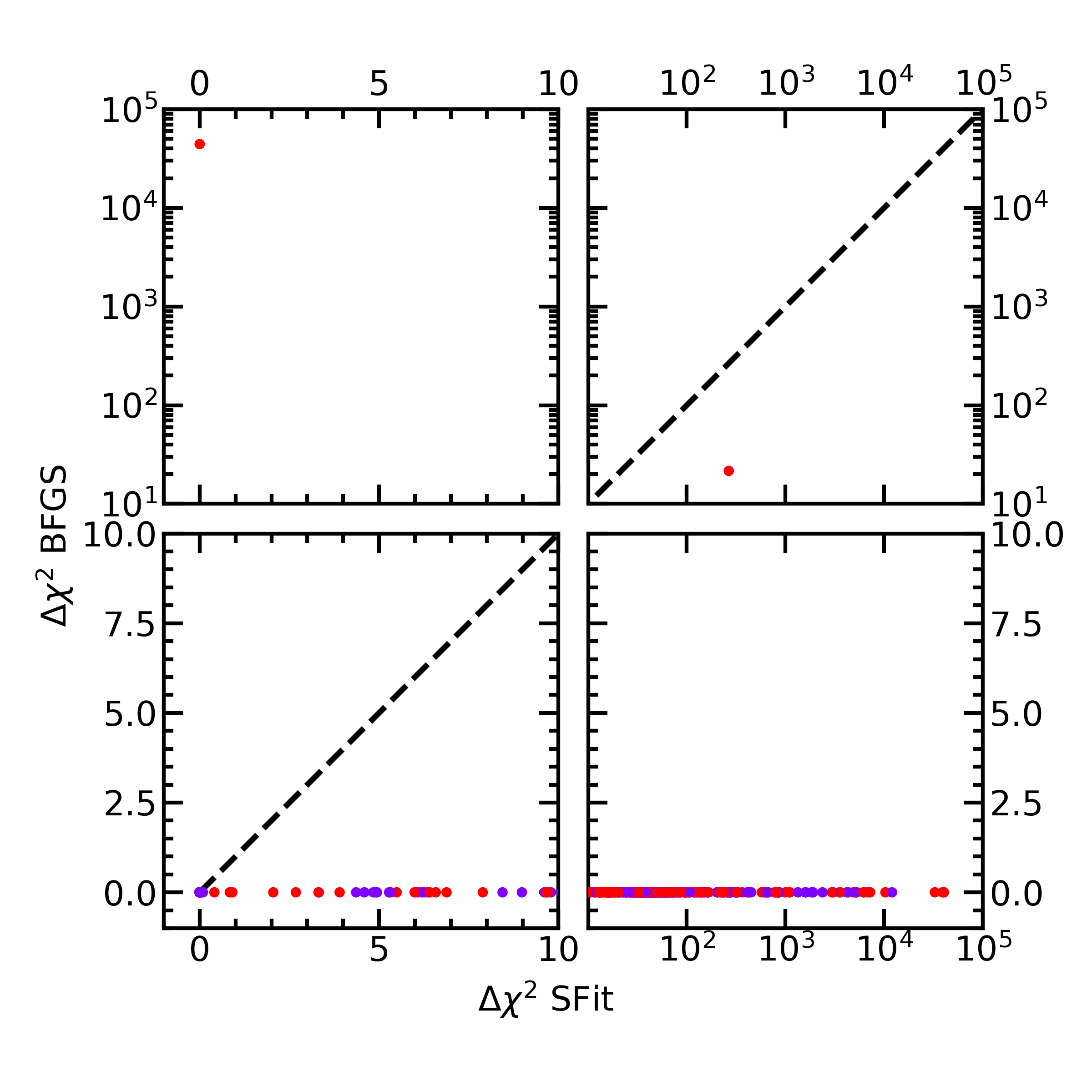}
	\caption{$\Delta\chi^2$ of the \bfgs\, model relative to the best-fitting model vs. $\Delta\chi^2$ of the \sfit\, model. Fits reported as successes by \bfgs are plotted in purple, while those reported as failures are shown in red. Events that were fit successfully by both algorithms appear at (0, 0). In each set of four panels, the axes are split so that [0, 10] is on a linear scale and [10, $10^5$] is on a log scale. The vertical bands of points at $x=0$ are fits that were successfully fit by \sfit\, but failed to be fit by \bfgs; purple points in those bands are false positives for \bfgs. The horizontal bands of points at $y=0$ are points for which \bfgs\, successfully found the minimum but \sfit\, did not; red points in those bands are false negatives.
	 \label{fig:dchi2_bfgs}}
	 \end{centering}
\end{figure}

\begin{figure}
	\begin{centering}
	\includegraphics[height=0.5\textheight]{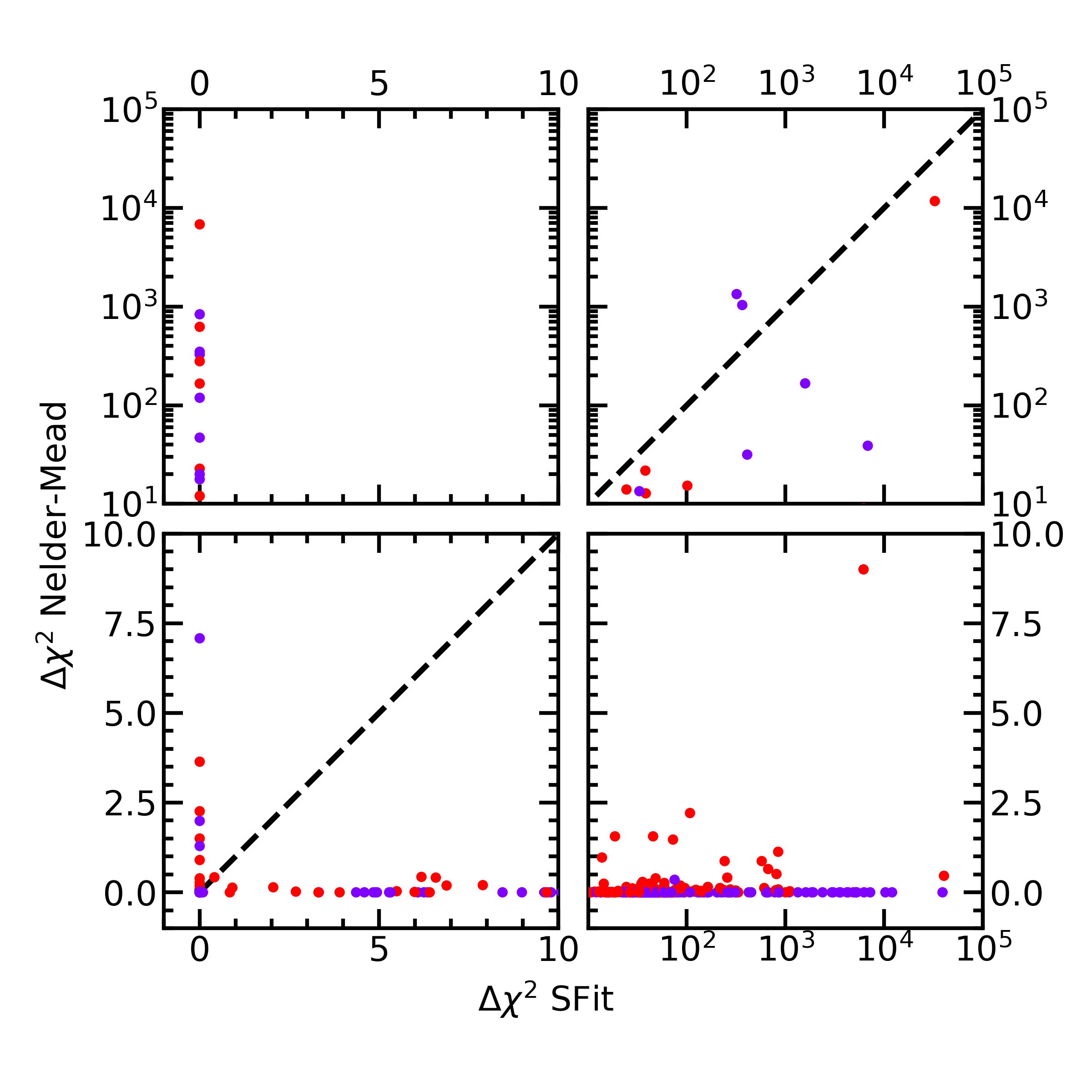}
	\caption{Same as Figure \ref{fig:dchi2_bfgs} but for the \neldermead\, algorithm relative to \sfit.
	 \label{fig:dchi2_nm}}
	 \end{centering}
\end{figure}

\begin{figure}
	\begin{centering}
	\includegraphics[height=0.5\textheight]{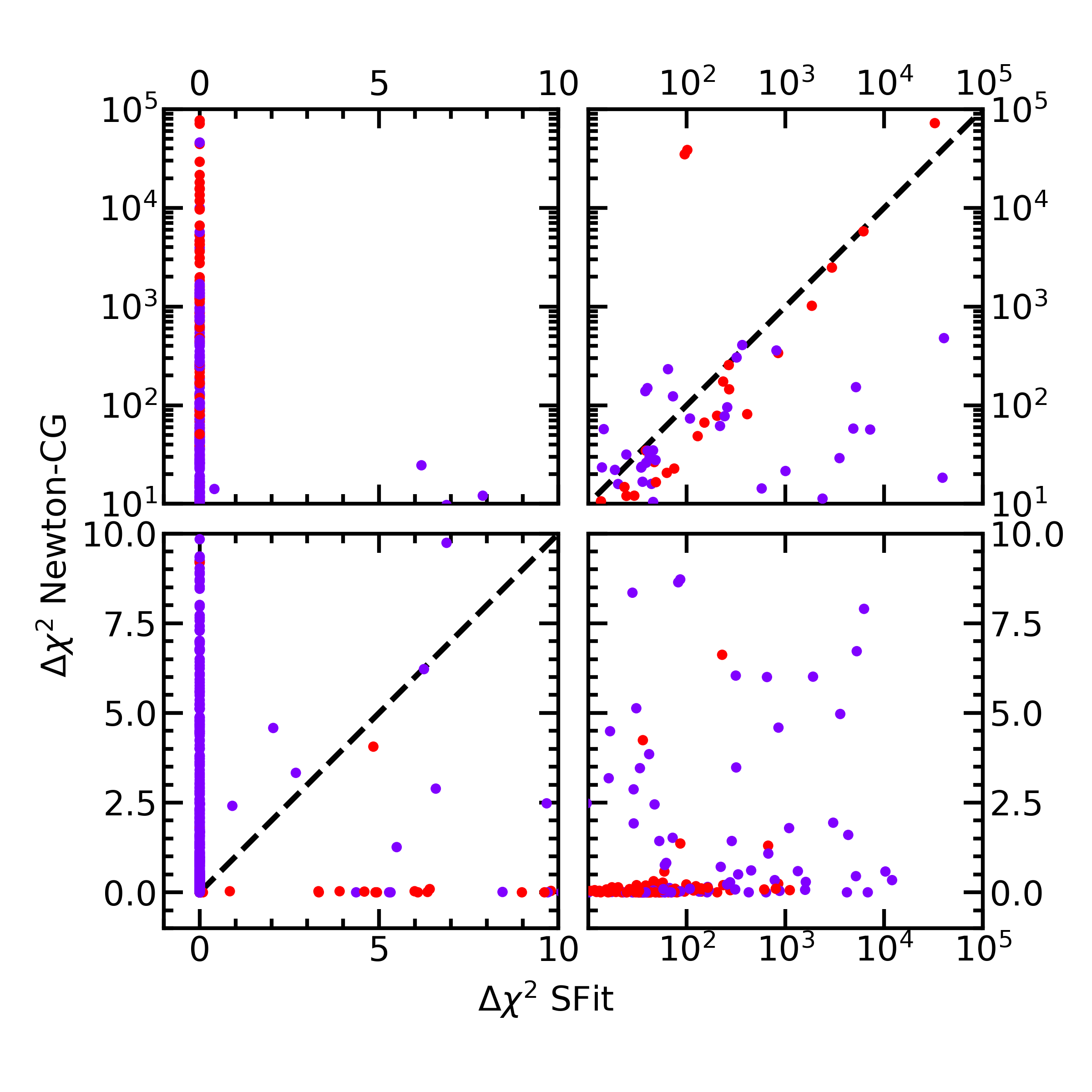}
	\caption{Same as Figure \ref{fig:dchi2_bfgs} but for the \newtoncg\, algorithm relative to \sfit.}
	 \label{fig:dchi2_newtoncg}
	 \end{centering}
\end{figure}

Figures \ref{fig:dchi2_bfgs}--\ref{fig:dchi2_newtoncg} compare the performance of \sfit\, to the other algorithms. All three plots have points that fall along the lines $x=0$ or $y=0$. This indicates that sometimes \sfit\, will successfully fit events that other algorithms do not and vice versa. For Figures \ref{fig:dchi2_nm} and \ref{fig:dchi2_newtoncg}, the mix of both colors (purple=reported successes and red=reported failures) along these lines also indicates that there is no category of \sfit\, fits (true positives, false positives, true negatives, false negatives) that is a subset of the same category for the other algorithm or vice versa. These qualitative impressions are quantified in the lower sections of Table \ref{tab:dchi2}.

In terms of number of $\chi^2$ function evaluations, \sfit\, is reasonably efficient. It scores in between the \bfgs\, and \neldermead\, algorithms (see Table \ref{tab:nfev}).

{\section{Summary}
\label{sec:summary}}

We presented an alternative method for generalizing Newton's method to minimize $\chi^2$ and its implementation as Python package called \sfit. We tested this implementation against the \bfgs, \neldermead, and \newtoncg\, algorithms in \minimize\, to objectively evaluate its performance in fitting point-lens microlensing light curves from the Korea Microlensing Telescope Network survey data. 

Of the three \minimize\, algorithms, \bfgs\, was able to find the best-fitting model almost 100\% of the time, despite reporting failed fits for 32\% of light curves. \newtoncg\, performed the worst with high rates of both false positives and false negatives. The \neldermead\, algorithm performed well, successfully finding the $\chi^2$ minimum for 98\% of light curves, but with a significant number of false negatives and requiring the most number of function evaluations.

We find that \sfit\, is able to successfully fit 83\% of point-lens microlensing light curves. It is characterized by high reliability, with extremely low rates of both false positives and false negatives. It is also relatively efficient, requiring a median of 167 function evaluations to meet the required tolerance. An additional advantage of this algorithm and implementation is that it automatically estimates uncertainties in the fitted parameters.

\vspace{12pt}
The Python implementation of this algorithm, including its specific application to microlensing events, can be found on \dataset[GitHub]{https://github.com/jenniferyee/sfit_minimizer}.

\section*{Acknowledgments}

We thank Radek Poleski and Keto Zhang for helpful discussions in the development of the code.
J.C.Y. acknowledges support from U.S. NSF Grant No. AST-2108414 and NASA Grant No. 22-RMAN22-0078. 
This research has made use of publicly available data 
(https://kmtnet.kasi.re.kr/ulens/) from the KMTNet system
operated by the Korea Astronomy and Space Science Institute
(KASI) at three host sites of CTIO in Chile, SAAO in South
Africa, and SSO in Australia. Data transfer from the host site to
KASI was supported by the Korea Research Environment
Open NETwork (KREONET).

\FloatBarrier
\newpage

\begin{deluxetable}{lr|rr|rr|rr|rr}
\tablecaption{Number of Fits with $\Delta\chi^2 < X$ of the Best-Fit}
\label{tab:dchi2}
\tablehead{
\colhead{} & \colhead{} & \multicolumn{4}{c}{$\Delta\chi^2 <$}\\
\colhead{Algorithm} & \colhead{Total} & \multicolumn{2}{|c}{0.1} & \multicolumn{2}{|c}{1.0} & \multicolumn{2}{|c}{10.0} & \multicolumn{2}{|c}{100.0} \\\colhead{} & \colhead{} & \multicolumn{1}{|c}{N} & \colhead{\%} & \multicolumn{1}{|c}{N} & \colhead{\%} & \multicolumn{1}{|c}{N} & \colhead{\%} & \multicolumn{1}{|c}{N} & \colhead{\%}
}
\startdata
\hline\hline
\multicolumn{10}{l}{All 1716 Events:}\\
\hline\hline
\multicolumn{10}{l}{Algorithm Reported Success:}\\
\bfgs                 & 1172 & 1172 & 100 & 1172 & 100 & 1172 & 100 & 1172 & 100 \\
\neldermead           & 1488 & 1470 &  99 & 1471 &  99 & 1474 &  99 & 1481 & 100 \\
\newtoncg             & 1307 &  590 &  45 &  843 &  64 & 1074 &  82 & 1225 &  94 \\
\sfit                 & 1425 & 1425 & 100 & 1425 & 100 & 1425 & 100 & 1425 & 100 \\
\hline
\multicolumn{10}{l}{Algorithm Reported Failure:}\\
\bfgs                 &  544 &  542 & 100 &  542 & 100 &  542 & 100 &  543 & 100 \\
\neldermead           &  228 &  170 &  75 &  208 &  91 &  217 &  95 &  223 &  98 \\
\newtoncg             &  409 &  295 &  72 &  318 &  78 &  325 &  79 &  351 &  86 \\
\sfit                 &  291 &    1 &   0 &    4 &   1 &   32 &  11 &  206 &  71 \\
\hline\hline
\multicolumn{10}{l}{1425 Events for which \sfit\, reported success:}\\
\hline\hline
\multicolumn{10}{l}{Algorithm Reported Success:}\\
\bfgs                 & 1038 & 1038 & 100 & 1038 & 100 & 1038 & 100 & 1038 & 100 \\
\neldermead           & 1345 & 1335 &  99 & 1335 &  99 & 1338 &  99 & 1341 & 100 \\
\newtoncg             & 1162 &  533 &  46 &  770 &  66 &  967 &  83 & 1089 &  94 \\
\hline
\multicolumn{10}{l}{Algorithm Reported Failure:}\\
\bfgs                 &  387 &  386 & 100 &  386 & 100 &  386 & 100 &  386 & 100 \\
\neldermead           &   80 &   67 &  84 &   71 &  89 &   74 &  92 &   76 &  95 \\
\newtoncg             &  263 &  199 &  76 &  201 &  76 &  203 &  77 &  216 &  82 \\
\hline\hline
\multicolumn{10}{l}{291 Events for which \sfit\, reported failure:}\\
\hline\hline
\multicolumn{10}{l}{Algorithm Reported Success:}\\
\bfgs                 &  134 &  134 & 100 &  134 & 100 &  134 & 100 &  134 & 100 \\
\neldermead           &  143 &  135 &  94 &  136 &  95 &  136 &  95 &  140 &  98 \\
\newtoncg             &  145 &   57 &  39 &   73 &  50 &  107 &  74 &  136 &  94 \\
\hline
\multicolumn{10}{l}{Algorithm Reported Failure:}\\
\bfgs                 &  157 &  156 &  99 &  156 &  99 &  156 &  99 &  157 & 100 \\
\neldermead           &  148 &  103 &  70 &  137 &  93 &  143 &  97 &  147 &  99 \\
\newtoncg             &  146 &   96 &  66 &  117 &  80 &  122 &  84 &  135 &  92 \\
\enddata
\end{deluxetable}

\begin{deluxetable}{lrrrr}
\tablecaption{Number of $\chi^2$ Function Evalutions}
\label{tab:nfev}
\tablehead{
\colhead{Algorithm} & \colhead{Mean} &  \colhead{Median} &  \colhead{StdDev} &  \colhead{Max}%\tablenotemark{a}}
}
\startdata
\bfgs                 &  94.1 &  34 & 191.5 & 814 \\
\neldermead           & 430.7 & 419 & 106.4 & 600 \\
\newtoncg             &  78.8 &  69 &  82.9 & 625 \\
\sfit                 & 175.8 & 167 & 115.1 & 583 \\
\enddata
%\tablenotetext{a}{Fitting terminated after 999 function evaluations.}
\end{deluxetable}

\end{document}